\RequirePackage{amsmath}
\documentclass[runningheads]{llncs}
\usepackage[T1]{fontenc}
\usepackage{graphicx}
\usepackage{amssymb}
\usepackage{graphicx}
\usepackage{subcaption}
\usepackage{placeins}
\usepackage{multirow}
\usepackage{booktabs}
\usepackage[hidelinks]{hyperref}
\begin{document}
	\emergencystretch 3em
	\title{Heuristic Modularity Maximization Algorithms for Community Detection Rarely Return an Optimal Partition or Anything Similar}
	\titlerunning{Evaluating Modularity-Based Community Detection Heuristics}
	\author{Samin Aref\inst{1}\orcidID{0000-0002-5870-9253}  \and \\
		Mahdi Mostajabdaveh\inst{2}\orcidID{0000-0002-2816-909X}
		\and \\
		Hriday Chheda\inst{3}\orcidID{0000-0002-9049-0255}}
	\authorrunning{S. Aref et al.}
	\institute{Department of Mechanical and Industrial Engineering, University of Toronto, M5S3G8, Canada \email{aref@mie.utoronto.ca}
		\and
		Huawei Technologies Canada, V5C 6S7, Canada \\
		\and
		Department of Computer Science, University of Toronto, M5S2E4, Canada
	}
	\maketitle              
	\begin{abstract}
		Community detection is a fundamental problem in computational sciences with extensive applications in various fields. The most commonly used methods are the algorithms designed to maximize modularity over different partitions of the network nodes. Using 80 real and random networks from a wide range of contexts, we investigate the extent to which current heuristic modularity maximization algorithms succeed in returning maximum-modularity (optimal) partitions. We evaluate (1) the ratio of the algorithms' output modularity to the maximum modularity for each input graph, and (2) the maximum similarity between their output partition and any optimal partition of that graph. We compare eight existing heuristic algorithms against an exact integer programming method that globally maximizes modularity. The average modularity-based heuristic algorithm returns optimal partitions for only 19.4\% of the 80 graphs considered.
		Additionally, results on adjusted mutual information reveal substantial dissimilarity between the sub-optimal partitions and any optimal partition of the networks in our experiments. More importantly, our results show that near-optimal partitions are often disproportionately dissimilar to any optimal partition. Taken together, our analysis points to a crucial limitation of commonly used modularity-based heuristics for discovering communities: they rarely produce an optimal partition or a partition resembling an optimal partition. If modularity is to be used for detecting communities, exact or approximate optimization algorithms are recommendable for a more methodologically sound usage of modularity within its applicability limits.
		\\
		
		This is a post-peer-review accepted manuscript from the Proceedings of the 23rd International Conference on Computational Science (ICCS 2023). The publisher authenticated version (version of record) is available on Springer-Nature website. \href{https://doi.org/10.1007/978-3-031-36027-5_48}{doi.org/10.1007/978-3-031-36027-5\_48}
		
	\end{abstract}
	%
	%
	%
	
	
	\section{Introduction}
	Community detection (CD), the process of inductively identifying communities within a network, is a core problem in computational sciences, particularly, in physics, computer science, biology, and computational social science \cite{zhang2014,fortunato2022newman}. Among common approaches for CD are the algorithms which are designed to maximize a utility function, modularity \cite{newman_modularity_2006}, across all possible ways that the nodes of the input network can be partitioned into communities. Modularity measures the fraction of edges within communities minus the expected fraction if the edges were distributed randomly; with the random distribution of the edges being a null model that preserves the node degrees. Despite their name and design philosophy, current modularity maximization algorithms, which are used by no less than tens of thousands of peer-reviewed studies \cite{Kosowski2020}, are not guaranteed to maximize modularity \cite{newman_equivalence_2016,kawamoto2019counting,meeks2020parameterised}. This has led to uncertainty \cite{good_performance_2010,kawamoto2019counting} in the extent to which they succeed in returning a maximum-modularity (optimal) partition or something similar.
	
	Modularity is among the first objective functions proposed for optimization-based detection of communities \cite{newman_modularity_2006,fortunato2016}.
	Several limitations \cite{Guimer2004,fortunato2010community,fortunato2016,peixoto_2023} of modularity including the resolution limit \cite{fortunato_2007} have led researchers to develop alternative CD methods using stochastic block modeling \cite{Karrer_2011,sbm_2014,liu2021scalable,serrano2021community}, information theoretic approaches \cite{rosvall_2007,rosvall_2008}, and alternative objective functions \cite{Aldecoa_2011_surprise,traag_2015_surprise,Miasnikof2020,marchese2022detecting}. Modularity-based algorithms are the most commonly used method for CD \cite{sobolevsky2014general,fortunato2022newman}. Despite the widespread adoption of modularity-based heuristics, there is uncertainty \cite{good_performance_2010,kawamoto2019counting} in their success in maximizing modularity. This study aims to address this uncertainty by quantifying the extent to which eight commonly used heuristics \cite{clauset_finding_2004,blondel_fast_2008,rb_pots_2008,sobolevsky2014general,zhang2014,paris_2018,traag_louvain_2019,edmot_2019} succeed in returning an optimal partition or a partition resembling an optimal partition. After describing the methods and materials, we present the main results followed by a discussion of the methodological ramifications and future directions.
	
	\section{Methods and Materials}
	This study aims to investigate the extent to which eight commonly used heuristic modularity maximization algorithms \cite{clauset_finding_2004,blondel_fast_2008,rb_pots_2008,sobolevsky2014general,zhang2014,paris_2018,traag_louvain_2019,edmot_2019} succeed in returning an optimal partition or a partition similar to an optimal partition. To achieve this objective, we quantify the proximity of their results to the globally optimal partition(s), which we obtain using an exact Integer Programming (IP) model for maximizing modularity \cite{brandes2007modularity,agarwal_modularity-maximizing_2008,dinh_toward_2015}. We do not claim that maximum-modularity partitions represent best partitions. Throughout the paper, we use the terms network and graph interchangeably.
	
	\subsection{Modularity}
	Consider the simple graph $G=(V,E)$ with $|V|=n$ nodes, $|E|=m$ edges, adjacency matrix entries $a_{ij}$, and a partition $X=\{V_1,V_2, \dots, V_k \}$ of the node set $V$ into $k$ communities. The modularity function $Q_{(G,X)}$ is computed \cite{newman_modularity_2006,fortunato2016} according to Eq.\ \eqref{eq0}
	
	\begin{equation}
		\label{eq0}
		Q_{(G,X)}= \frac{1}{2m} \sum \limits_{(i,j) \in V^2} \left( a_{ij} - \gamma\frac{d_id_j}{2m}\right) \delta(i,j)
	\end{equation}
	
	\noindent where $d_i$ represents the degree of node $i$, $\gamma$ is the resolution parameter\footnote{Without loss of generality, we set $\gamma=1$ for all the analysis in this paper.}, and $\delta(i,j)$ is 1 if nodes $i$ and $j$ are in the same community otherwise 0. The term associated with each pair of nodes $(i,j)$ is alternatively represented as $b_{ij}=a_{ij} -\gamma\frac{d_id_j}{2m}$ and referred to as the modularity matrix entry for $(i,j)$. 
	
	\subsection{Modularity maximization}
	The modularity maximization problem for input graph $G=(V,E)$ involves finding a partition $X^*$ whose associated $Q_{(G,X^*)}$ is globally maximum over all possible partitions of the node set $V$.
	
	\subsection{Sparse IP formulation of modularity maximization}
	Consider the simple graph $G=(V,E)$ with modularity matrix entries $b_{ij}$, obtained using the resolution parameter $\gamma$. We use the binary decision variable $x_{ij}$ for each pair of distinct nodes $(i,j),i<j$. Their community membership is either the same (represented by $x_{ij}=0$) or different (represented by $x_{ij}=1$). Accordingly, the problem of maximizing the modularity of input graph $G$ can be formulated as an IP model \cite{dinh_toward_2015} as in Eq.\ \eqref{eq1}. 
	
	\begin{equation}
		\label{eq1}
		\begin{split}
			&\max_{x_{ij}} Q = \frac{1}{2m}   \left( \sum\limits_{(i,j) \in V^2 , i< j} 2b_{ij}(1- x_{ij}) + \sum\limits_{(i,i) \in V^2} b_{ii} \right) \\
			&\text{s.t.}  \quad  x_{ik}+x_{jk} \geq x_{ij} \quad \forall (i,j) \in V^2 , i< j, k\in K(i,j) \\ 
			& \quad \quad x_{ij} \in \{0,1\} \quad \forall (i,j) \in V^2 , i< j
		\end{split}
	\end{equation}
	
	In Eq.\ \eqref{eq1}, the optimal objective function value equals the maximum modularity for the input graph $G$. An optimal community assignment is characterized by the optimal values of the $x_{ij}$ variables. $K(i,j)$ indicates a minimum-cardinality separating set \cite{dinh_toward_2015} for the nodes $i,j$. Using $K(i,j)$ in the IP model of this problem leads to a more efficient formulation with $\mathcal{O}(n^2)$ constraints \cite{dinh_toward_2015} instead of $\mathcal{O}(n^3)$ constraints as in earlier IP formulations of the problem \cite{brandes2007modularity,agarwal_modularity-maximizing_2008}.
	Solving this optimization problem is NP-complete \cite{brandes2007modularity,meeks2020parameterised}. We use the \textit{Gurobi} solver (version 10.0) \cite{gurobi} to solve it for the small and mid-sized instances as outlined in Subsection \ref{ss:data}. 
	
	\subsection{Reviewing eight heuristic modularity maximization algorithms}
	
	We evaluate eight modularity maximization heuristics known as Clauset-Newman-Moore (CNM) \cite{clauset_finding_2004}, Louvain \cite{blondel_fast_2008}, Leicht-Newman (LN) \cite{rb_pots_2008}, Combo \cite{sobolevsky2014general}, Belief \cite{zhang2014}, Paris \cite{paris_2018}, Leiden \cite{traag_louvain_2019}, and EdMot-Louvain \cite{edmot_2019}. We have used the Python implementations of these eight algorithms which are accessible in the Community Discovery library (\textit{CDlib}) version 0.2.6 \cite{rossetti2019cdlib}. 
	
	We briefly describe how these eight algorithms use modularity to discover communities.
	The CNM algorithm initializes each node as a community by itself. It then follows a greedy scheme of merging two communities that contribute the maximum positive value to modularity \cite{clauset_finding_2004}.
	The Louvain algorithm involves two sets of iterative steps: (1) locally moving nodes for increasing modularity and (2) aggregating the communities from the first step \cite{blondel_fast_2008}. Despite Louvain being the most commonly used modularity-based algorithm \cite{Kosowski2020}, it may sometimes lead to disconnected components in the same community \cite{traag_louvain_2019}.
	The LN algorithm uses spectral optimization to maximize modularity which also supports directed graphs \cite{rb_pots_2008}.
	The Combo algorithm is a general optimization-based CD method which supports modularity maximization among other tasks. It involves two sets of iterative steps: (1) finding the best merger, split, or recombination of communities to maximize modularity and (2) performing a series of Kernighan-Lin bisections \cite{kernighan1970efficient} on the communities as long as they increase modularity \cite{sobolevsky2014general}.
	The Belief algorithm seeks the consensus of different high-modularity partitions through a message-passing algorithm \cite{zhang2014} motivated by the premise that maximizing modularity can lead to many poorly correlated competing partitions.
	The Paris algorithm is suggested to be a modularity-maximization scheme with a sliding resolution \cite{paris_2018}; that is, an algorithm capable of capturing the multi-scale community structure of real networks without a resolution parameter. It generates a hierarchical community structure based on a simple distance between communities using a nearest-neighbour chain \cite{paris_2018}.
	The Leiden algorithm attempts to resolve a defect of the Louvain algorithm in returning badly connected communities. It is suggested to guarantee well-connected communities in which all subsets of all communities are locally optimally assigned \cite{traag_louvain_2019}. The EdMot-Louvain algorithm (EdMot for short) is developed to overcome the hypergraph fragmentation issue observed in previous motif-based CD methods \cite{edmot_2019}. It first creates the graph of higher-order motifs (small dense subgraph patterns) and then partitions it using the Louvain method to heuristically maximize modularity using higher-order motifs \cite{edmot_2019}.
	
	To evaluate these eight modularity-based algorithms in maximizing modularity, we quantify (1) the ratio of their output modularity to the maximum modularity for each input graph and (2) the maximum similarity between their output partition and any optimal partition of that graph. We obtain optimal partitions by solving the IP model in Eq.\ \eqref{eq1} using the Gurobi solver (version 10.0) with a termination criterion ensuring global optimality \cite{gurobi}.

	\subsection{Measures for evaluating heuristic algorithms}
	
	For a quantitative measure of proximity to global optimality, we define and use the \textit{Global Optimality Percentage} (GOP) as the fraction of the modularity returned by a heuristic method for a network divided by the globally maximum modularity for that network (obtained by solving the IP model in Eq.\ \eqref{eq1}). In all cases where the modularity returned by a heuristic method equals the maximum modularity for the input graph, we set GOP=1. In cases where a heuristic algorithm returns a partition with a negative modularity value, we set GOP=0 to facilitate easier interpretation of proximity to optimality based on non-negative GOP values. 
	
	We also use a quantitative measure for the similarity of a partition to an optimal partition. Normalized Adjusted Mutual Information (AMI) \cite{vinh_AMI} is a measure of similarity between two partitions of the same network. Unlike normalized mutual information \cite{vinh_AMI}, AMI adjusts the measurement based on the similarity that two partitions may have by pure chance. AMI for a pair of identical partitions (or permutations of the same partition) equals 1. For two different partitions, however, AMI takes a smaller value (including 0 or negative values close to 0 for two extremely dissimilar partitions). 
	
	\subsection{Data and resources}
	\label{ss:data}
	
	For our computational experiments, we include {60 real networks}\footnote{All networks are accessible from the \href{https://networks.skewed.de/}{Netzschleuder} with the details in the Appendix.} with no more than {2812 edges} as well as 10 Erd\H{o}s-R\'{e}nyi graphs and 10 Barab\'{a}si-Albert graphs with 125-153 edges. These instance sizes were chosen to ensure all algorithms terminate within a reasonable time. The computational experiments were implemented in Python 3.9 using a notebook computer with an Intel Core i7-11800H @ 2.30GHz CPU and 64 GB of RAM running Windows 10.
	
	\section{Results}
	
	We present the main results from our experiments in the following four subsections. In Subsection \ref{sec:results1}, we compare partitions from different algorithms on a single network. In Subsection \ref{sec:results2}, we examine the multiplicity of optimal partitions and investigate the similarity between multiple optimal partitions of the same networks. In Subsection \ref{sec:results3}, we evaluate the effectiveness of the heuristic algorithms on 80 networks by measuring the distance of sub-optimal partitions from an optimal partition. Finally, in Subsection \ref{sec:results4}, we investigate the success rate of the heuristic algorithms in finding an optimal partition.
	
	\subsection{Comparing partitions from different algorithms on one network}\label{sec:results1}
\begin{figure}[!htb]
		\centering
		\begin{subfigure}[b]{0.32\textwidth}
			\centering
			\includegraphics[trim={10.6cm 10.6cm 10.6cm 10.6cm},clip,width=\textwidth]{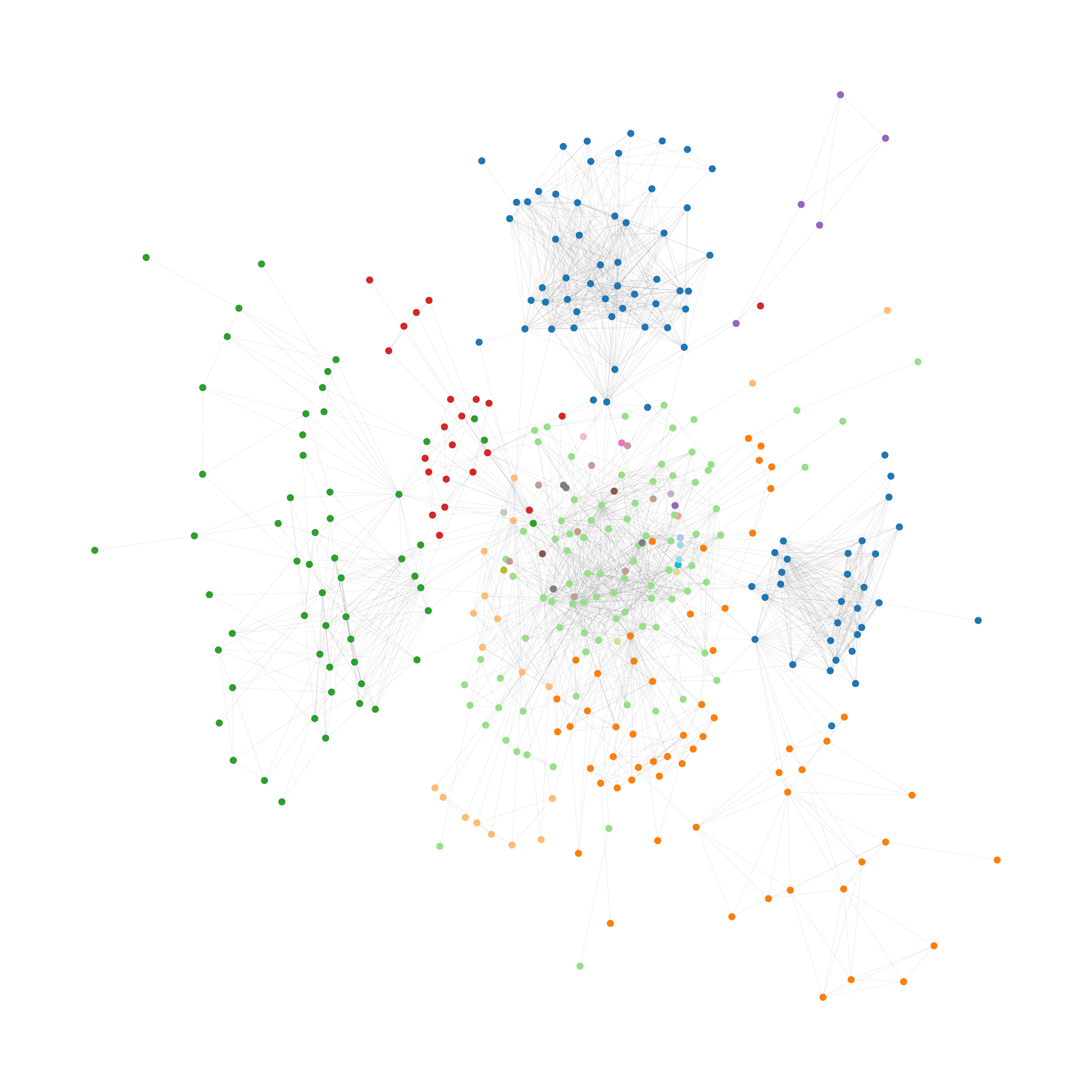}
			\caption{IP, $Q^*=0.7157714,
				\\\hspace{\textwidth} k=28, \text{AMI}=1$}
			\label{subfig:ip}
		\end{subfigure}
		\begin{subfigure}[b]{0.32\textwidth}
			\centering
			\includegraphics[trim={10.6cm 10.6cm 10.6cm 10.6cm},clip,width=\textwidth]{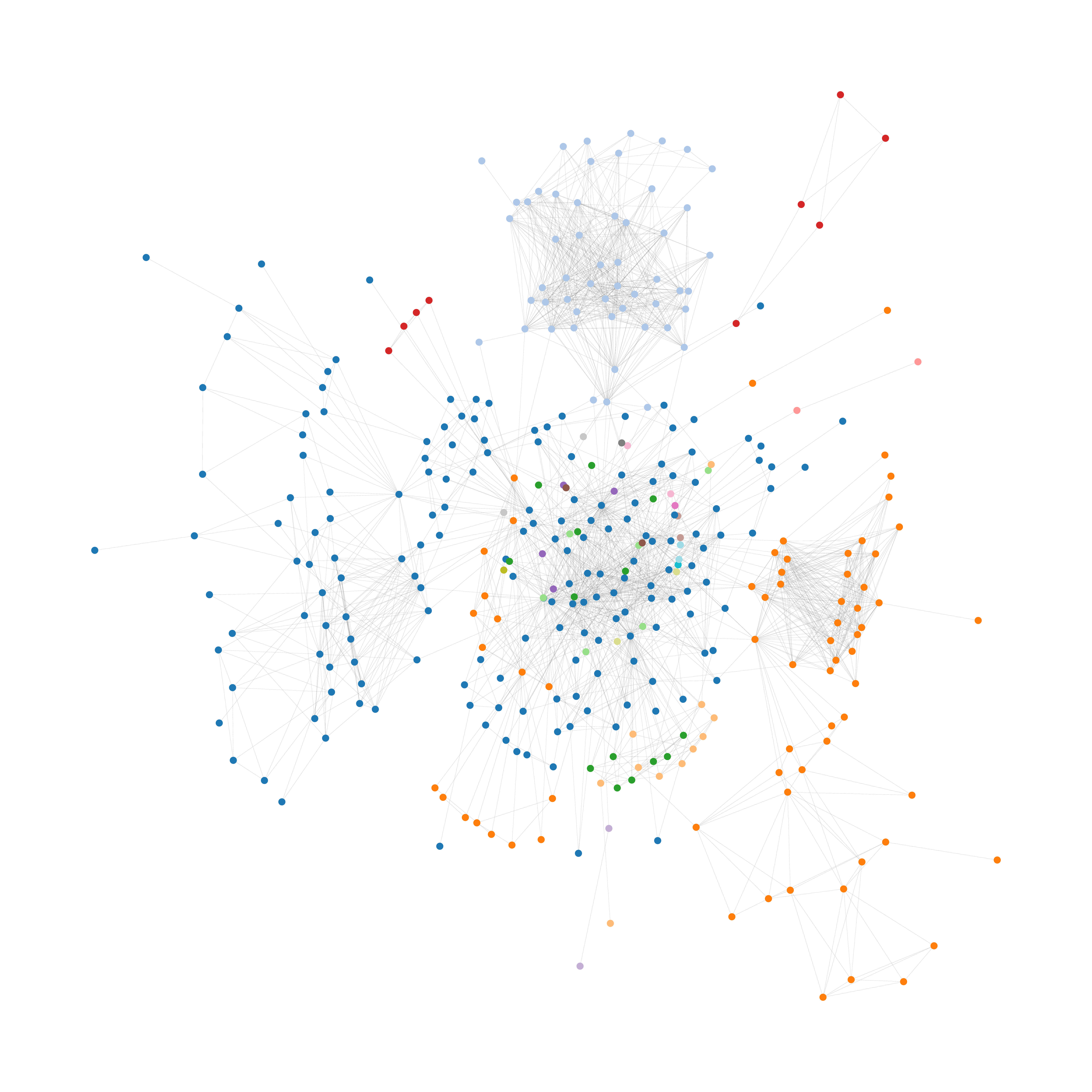}
			\caption{CNM, $Q=0.6971,\\\hspace{\textwidth} k=30, \text{AMI}=0.829$}
			\label{subfig:cnm}
		\end{subfigure}
		\begin{subfigure}[b]{0.32\textwidth}
			\centering
			\includegraphics[trim={10.6cm 10.6cm 10.6cm 10.6cm},clip,width=\textwidth]{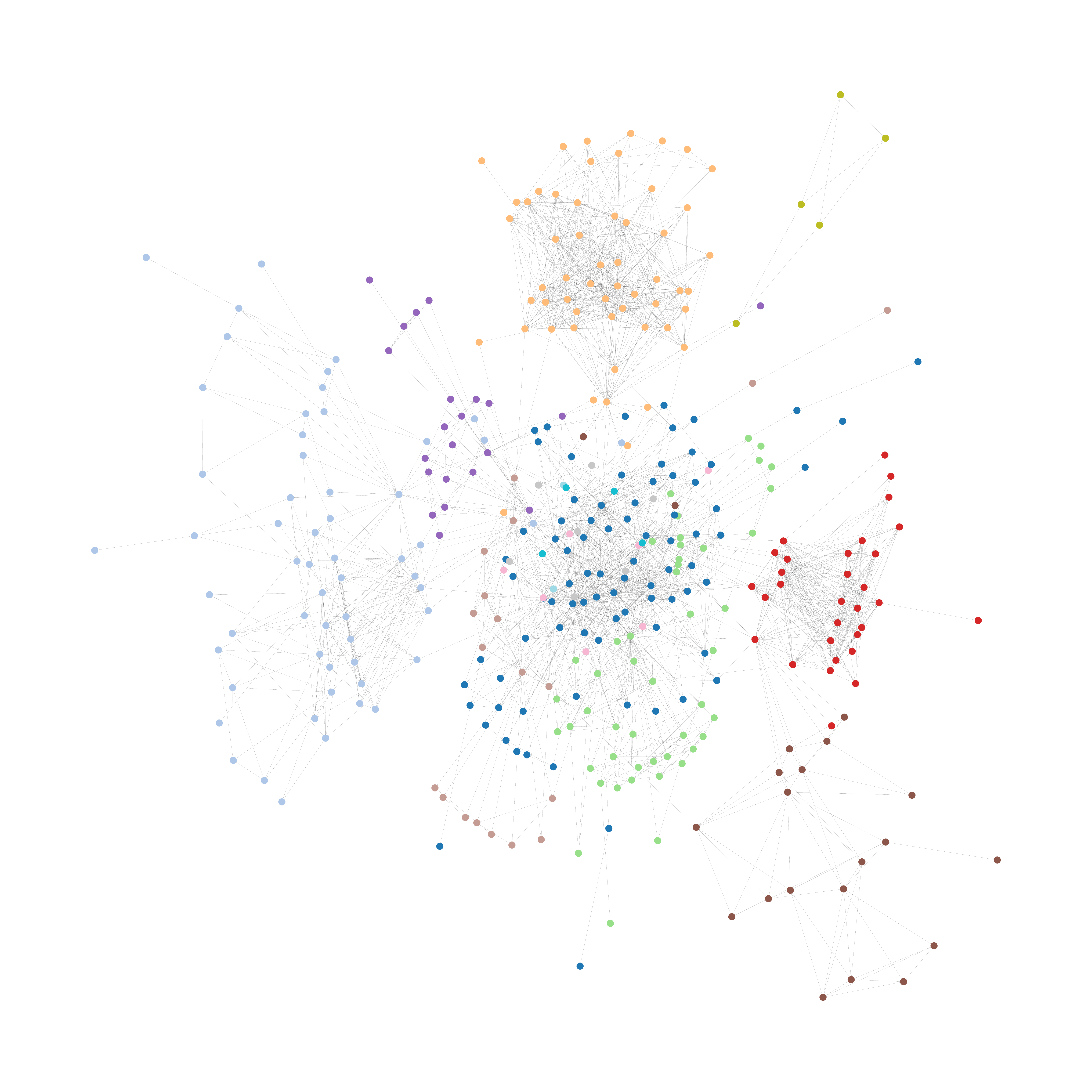}
			\caption{Combo, $Q=0.7157709,\\\hspace{\textwidth} k=13, \text{AMI}=0.949$}
			\label{subfig:combo}
		\end{subfigure}
		\begin{subfigure}[b]{0.32\textwidth}
			\centering
			\includegraphics[trim={10.6cm 10.6cm 10.6cm 10.6cm},clip,width=\textwidth]{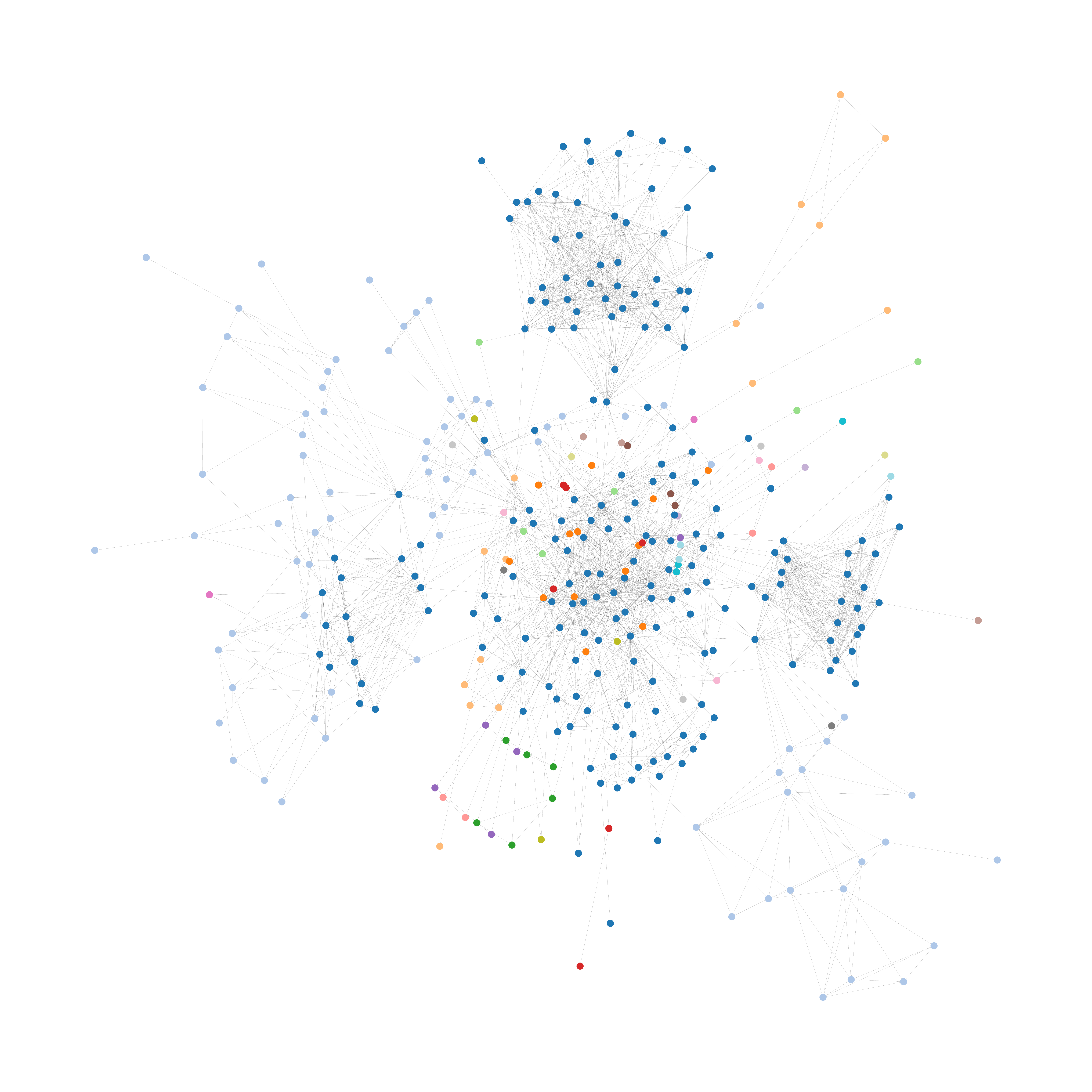}
			\caption{EdMot, $Q=0.4902,\\\hspace{\textwidth} k=53, \text{AMI}=0.651$}
			\label{subfig:edmot}
		\end{subfigure}
		\begin{subfigure}[b]{0.32\textwidth}
			\centering
			\includegraphics[trim={10.6cm 10.6cm 10.6cm 10.6cm},clip,width=\textwidth]{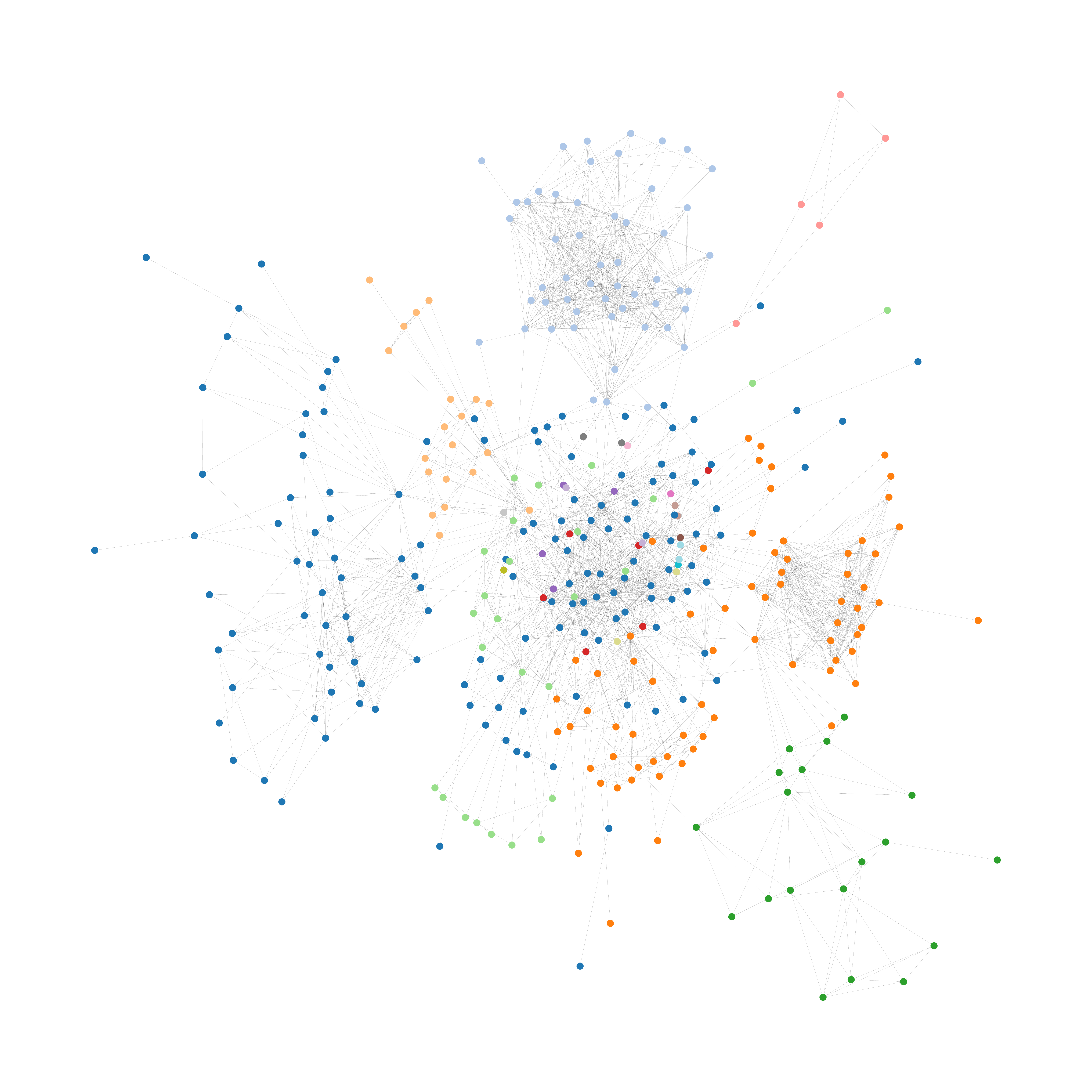}
			\caption{Leiden, $Q=0.7082,\\\hspace{\textwidth} k=32, \text{AMI}=0.908$}
			\label{subfig:leiden}
		\end{subfigure}
		\begin{subfigure}[b]{0.32\textwidth}
			\centering
			\includegraphics[trim={10.6cm 10.6cm 10.6cm 10.6cm},clip,width=\textwidth]{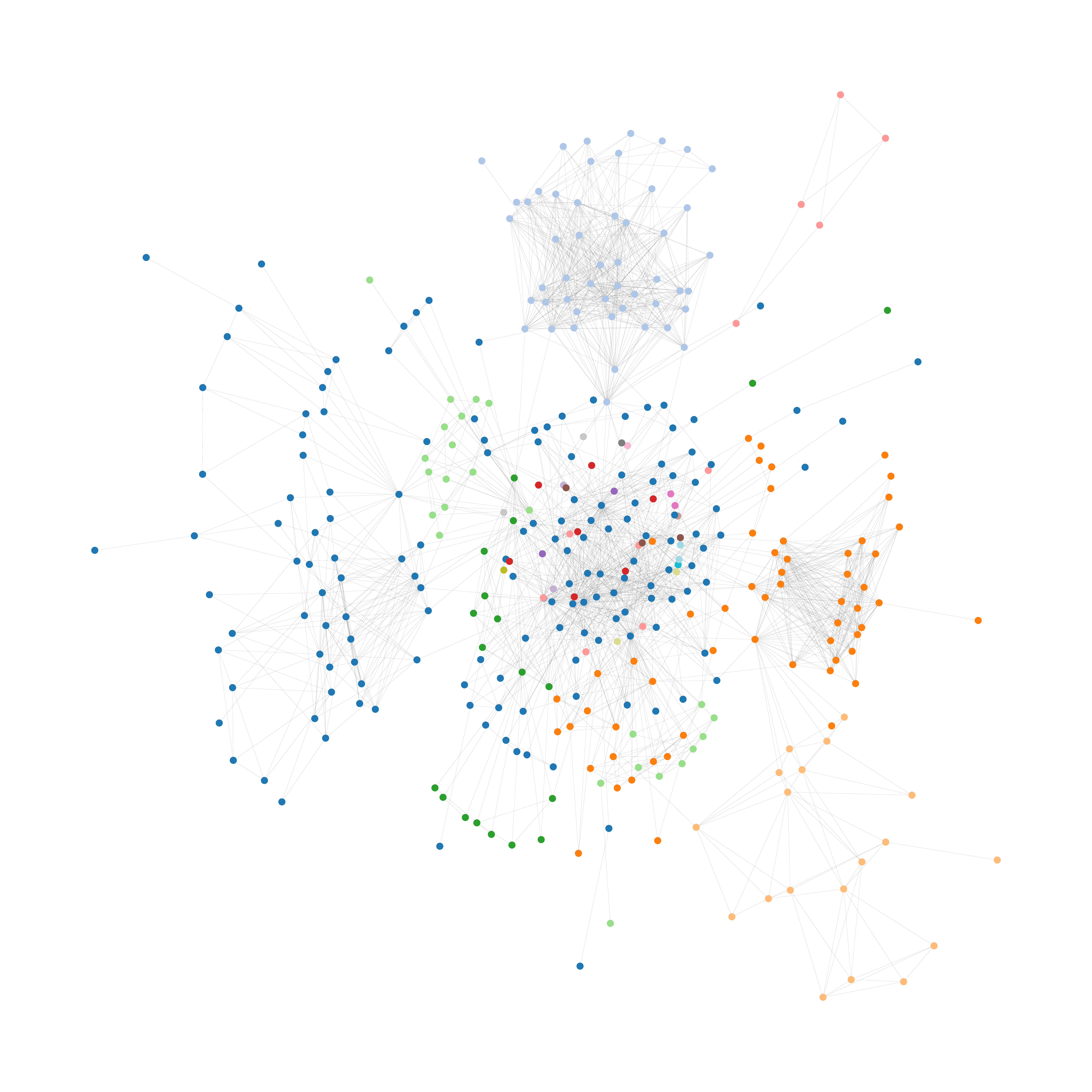}
			\caption{Louvain, $Q=0.7087,\\\hspace{\textwidth} k=29, \text{AMI}=0.920$}
			\label{subfig:louvain}
		\end{subfigure}
		\begin{subfigure}[b]{0.32\textwidth}
			\centering
			\includegraphics[trim={10.6cm 10.6cm 10.6cm 10.6cm},clip,width=\textwidth]{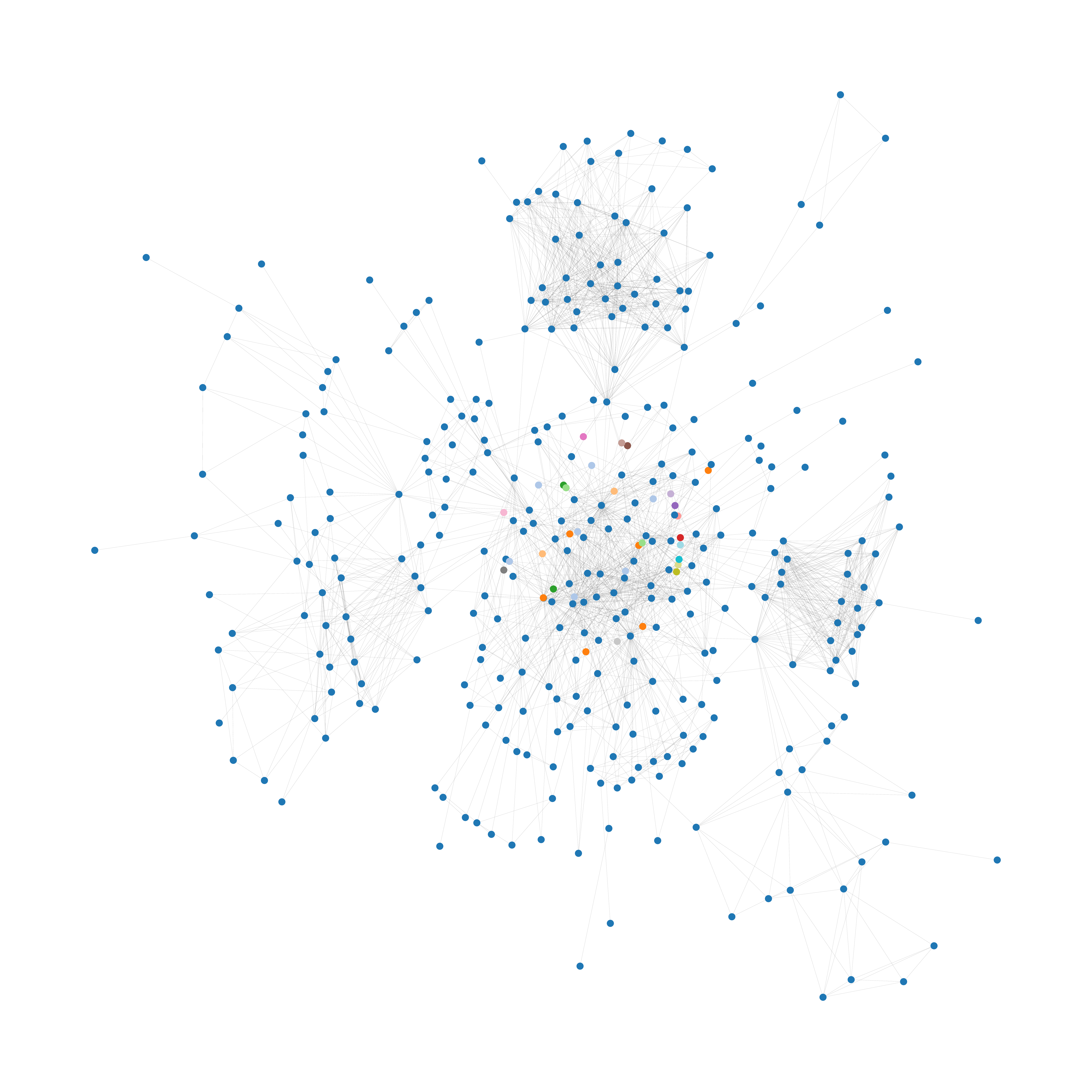}
			\caption{Paris, $Q=0.0338,\\\hspace{\textwidth} k=20, \text{AMI}=0.363$}
			\label{subfig:paris}
		\end{subfigure}
		\begin{subfigure}[b]{0.32\textwidth}
			\centering
			\includegraphics[trim={10.6cm 10.6cm 10.6cm 10.6cm},clip,width=\textwidth]{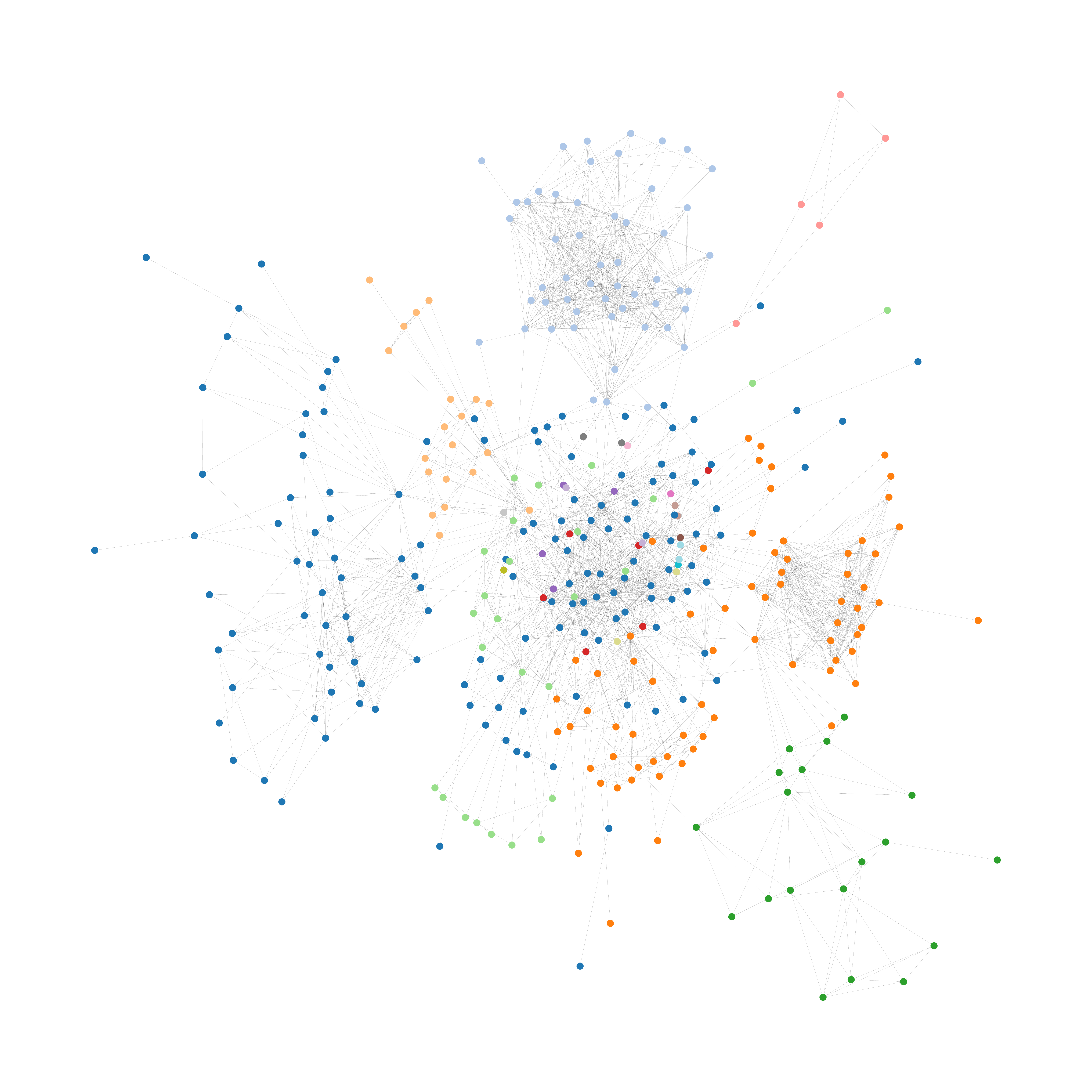}
			\caption{LN, $Q=0.7139,\\\hspace{\textwidth} k=28, \text{AMI}=0.971$}
			\label{subfig:ln}
		\end{subfigure}
		\begin{subfigure}[b]{0.32\textwidth}
			\centering
			\includegraphics[trim={10.6cm 10.6cm 10.6cm 10.6cm},clip,width=\textwidth]{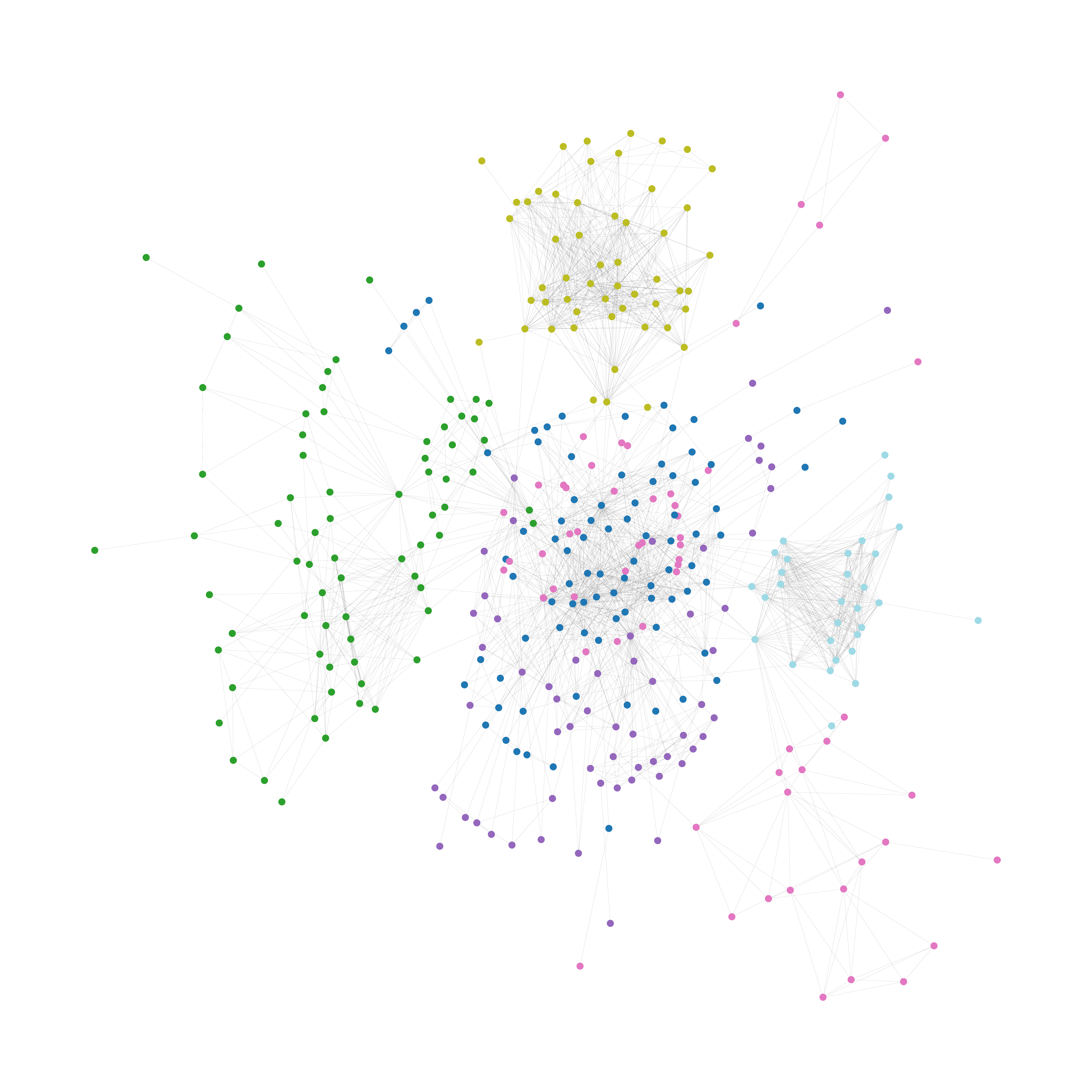}
			\caption{Belief, $Q=0.4566, \\\hspace{\textwidth}k=3, \text{AMI}=0.786$}
			\label{subfig:belief}
		\end{subfigure}
		\caption{Modularity maximization for one network using nine methods leading to one optimal partition (panel a) and eight sub-optimal partitions (panels b-i) with different $Q$, $k$, and AMI values. (Magnify the high-resolution color figure on screen for more details.) }
		\label{fig:facebook}
	\end{figure}

	Figure \ref{fig:facebook} shows one graph and its nine partitions returned by nine CD methods. This graph\footnote{ {facebook\_friends} network \cite{maier2017cover} from the \href{https://networks.skewed.de/}{Netzschleuder} repository} represents an anonymized Facebook \textit{ego network}\footnote{A network of one person's social ties to other persons and their ties to each other}. Nodes represent Facebook users, and an edge exists between any pair of users who were friends on Facebook in April 2014 \cite{maier2017cover}. Communities are shown using node colors.
	
	Panel \ref{subfig:ip} of Figure \ref{fig:facebook} shows an optimal partition obtained by solving the IP model in Eq.\ \eqref{eq1} for the network {facebook\_friends}. It involves $k=28$ communities, and a maximum modularity value of $Q^*=0.7157714$. The partitions from the eight heuristic modularity maximization algorithms are all sub-optimal as depicted in panels \ref{subfig:cnm}--\ref{subfig:belief} of Figure \ref{fig:facebook}. Compared to other algorithms, the two algorithms Combo and LN have more success in achieving proximity to an optimal partition. LN returns a partition with $k=28$ communities and a modularity of $Q=0.7139$ which has the highest AMI among all heuristics ($0.971$). The relative success of the Combo algorithm is in returning a high-modularity partition with $Q=0.7157709$, but with $k=13$ communities and a lower AMI ($0.949$) compared to LN. The sub-optimal partitions from the other six algorithms have more substantial variations in $Q$, AMI, and $k$ (number of communities) as shown by the values in the corresponding subcaptions in Figure \ref{fig:facebook}. 
	
	\subsection{Multiplicity of optimal partitions}\label{sec:results2}
	
\begin{figure}[!htb]
		\centering
		\begin{subfigure}[b]{0.48\textwidth}
			\centering
			\includegraphics[trim={8cm 8cm 8cm 8cm},clip,width=\textwidth]{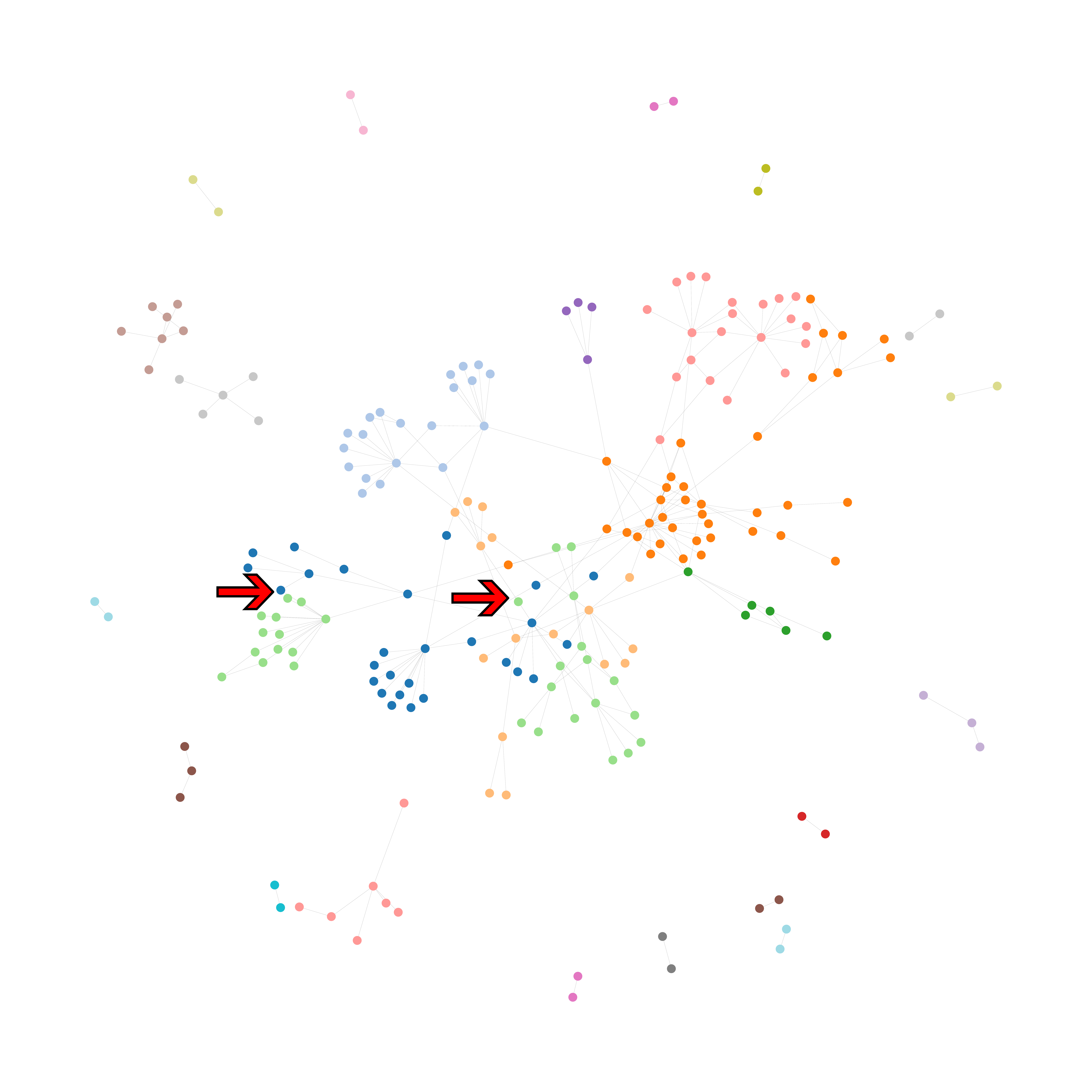}
			\caption{Indicated nodes are blue and green}
			\label{subfig:op1}
		\end{subfigure}
		\begin{subfigure}[b]{0.48\textwidth}
			\centering
			\includegraphics[trim={8cm 8cm 8cm 8cm},clip,width=\textwidth]{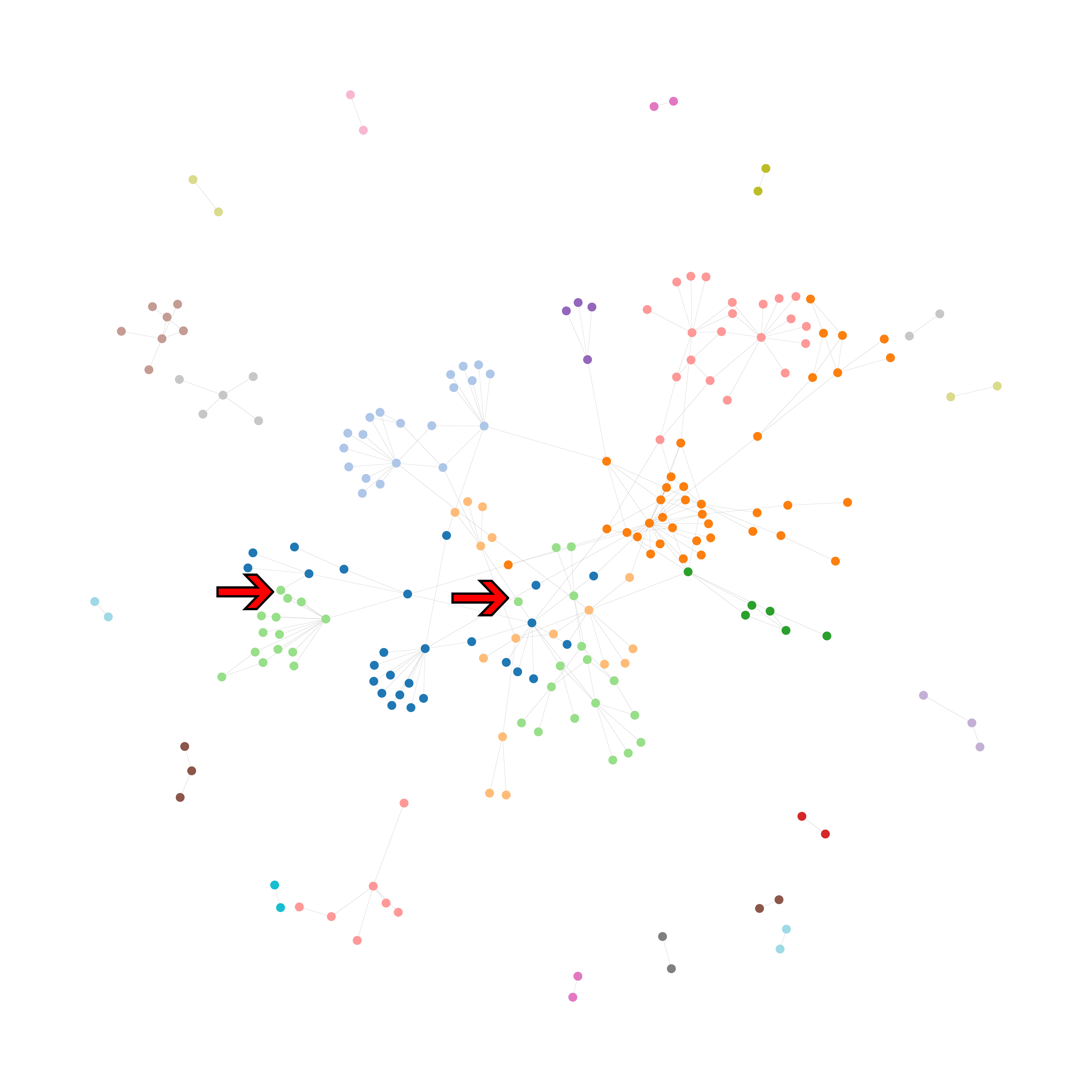}
			\caption{Indicated nodes are green and green}
			\label{subfig:op2}
		\end{subfigure}
		\centering
		\begin{subfigure}[b]{0.48\textwidth}
			\centering
			\includegraphics[trim={8cm 8cm 8cm 8cm},clip,width=\textwidth]{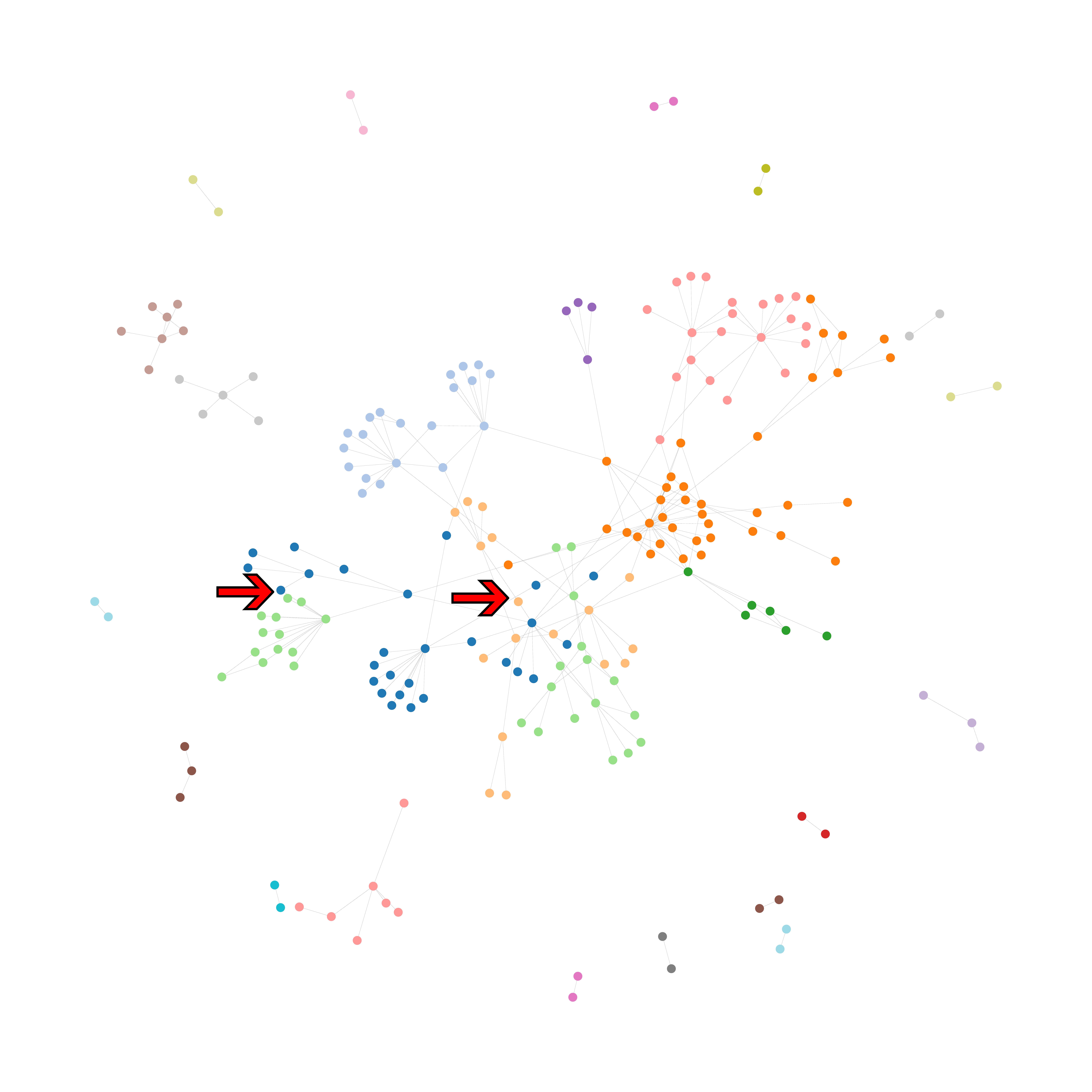}
			\caption{Indicated nodes are blue and orange}
			\label{subfig:op3}
		\end{subfigure}
		\begin{subfigure}[b]{0.48\textwidth}
			\centering
			\includegraphics[trim={8cm 8cm 8cm 8cm},clip,width=\textwidth]{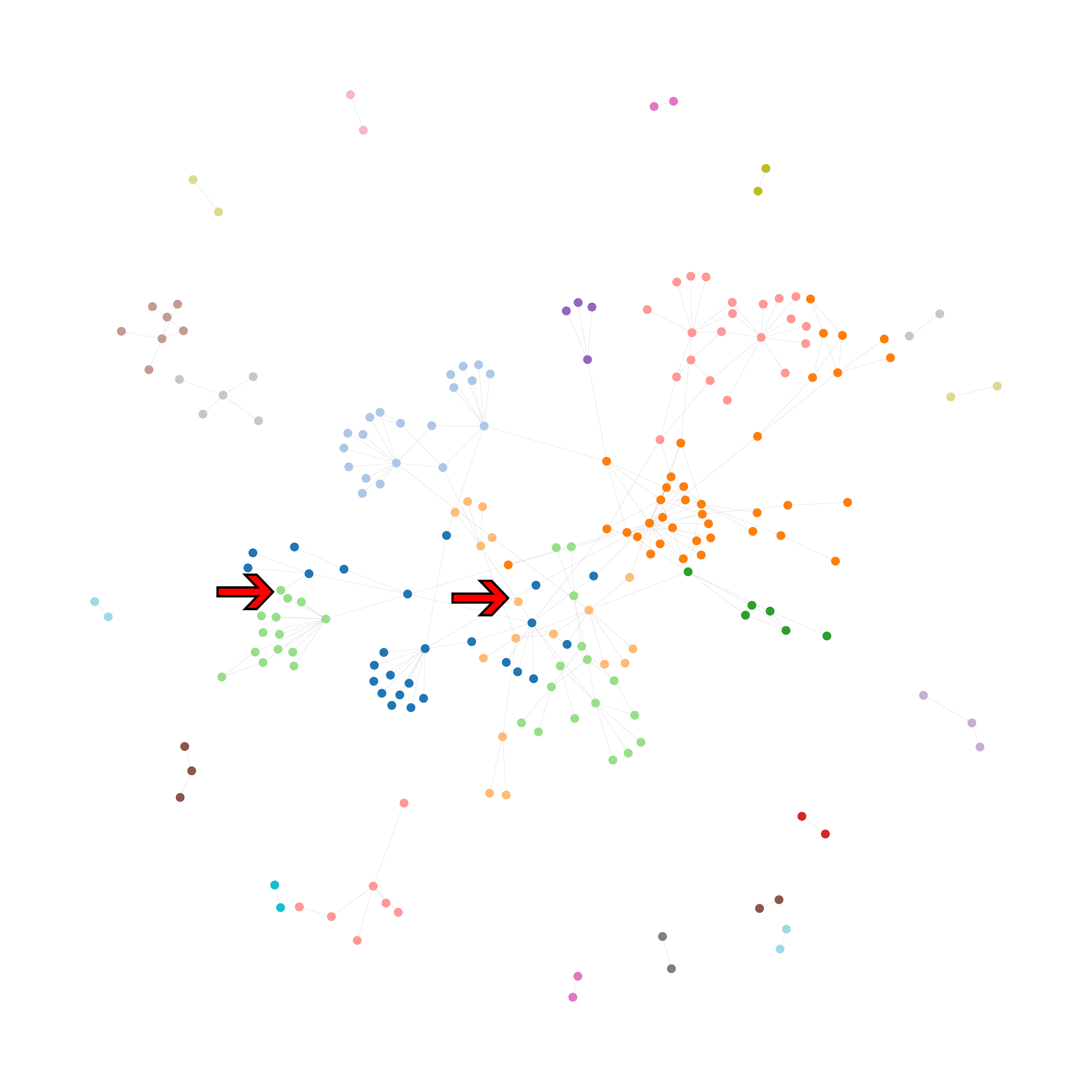}
			\caption{Indicated nodes are green and orange}
			\label{subfig:op4}
		\end{subfigure}
		\caption{A protein network and its four optimal partitions (panels a-d). The red arrows show the differences between optimal partitions. (Magnify the high-resolution color figure on screen for more details.)}
		\label{fig:multiplicity}
	\end{figure}

	While the partition which maximizes modularity is often unique, some graphs have multiple optimal partitions. For all networks considered in our analysis, we obtain all optimal partitions using the Gurobi solver by running it with a special configuration for finding all optimal partitions \cite{gurobi}. Figure \ref{fig:multiplicity} shows a protein network\footnote{ {interactome\_pdz} network \cite{pdzbase2005} from the \href{https://networks.skewed.de/}{Netzschleuder} repository} and its four optimal partitions. In this network, nodes represent proteins and an edge represents a binding interaction between two proteins (PDZ-domain-mediated protein–protein binding interaction) \cite{pdzbase2005}. All four optimal partitions have $Q^*=0.80267$ and $k=29$.
	
	The differences between optimal partitions of this network are in the community assignments for two nodes indicated by red arrows in Figure \ref{fig:multiplicity}. The six pairwise AMI values for the optimal partitions are all $>0.98$ confirming the high level of similarity between the four optimal partitions in Figure \ref{fig:multiplicity}.
	
	Obtaining all optimal partitions for all 80 networks, we observed that 89\% of the graphs have unique optimal partitions and the multiplicity of optimal partitions is a relatively rare event. Given the possibility of multiple optimal partitions in some graphs, we calculated the AMI for the partition of each heuristic algorithm and each of the multiple optimal partitions of that graph. We then conservatively reported the maximum AMI of each heuristic for each graph to quantify the similarity between that partition and its closest optimal partition. Consequently, a low value of AMI for a partition obtained by a heuristic algorithm indicates its dissimilarity to any optimal partition.
	
	Our results suggest that the rarely observed multiple optimal partitions of a graph often have a high degree of similarity (AMI values $>0.9$) because their differences are often only in the community assignments of a very few nodes (as in Figure \ref{fig:multiplicity}). Dissimilarity between multiple optimal partitions of a network seems to be exceptional, but it has been observed in one of our 80 networks: \textit{contiguous USA}\footnote{ {contiguous\_usa} network \cite{knuth1993stanford} from the \href{https://networks.skewed.de/}{Netzschleuder} repository}, where nodes are US states and each edge indicates a land-based border between two states. The AMI of the two optimal partitions for this network is exceptionally low $(0.34)$. Upon further investigation, we observed that one optimal partition combines five communities of the other optimal partition together. This makes the two partitions related in terms of belonging to a clustering hierarchy, while they are not similar according to an AMI definition of partition similarity. These exceptional cases are possible due to the mathematical symmetries resulted from the value of $\gamma$ used in Eq.\ \eqref{eq0} for defining modularity.
	Our results suggest that there is usually a distinct uniqueness to an optimal partition (or a group of similar optimal partitions) for a given network in comparison to sub-optimal partitions. This new perspective is contrary to the premise that maximizing modularity leads to many competing partitions with almost the same modularity \cite{zhang2014} and no clear way of selecting between them \cite{peixoto_2023}. It is the failure to actually maximize modularity that may lead to many poorly correlated competing partitions with unknown distances from the desired objective (both in modularity and in partition similarity). What remains to be analyzed is how different sub-optimal partitions are from an optimal partition and how often heuristic modularity maximization algorithms return sub-optimal partitions. We investigate these two questions in the next two subsections. 
	
	\subsection{Evaluating heuristic algorithms on 80 networks}\label{sec:results3}
	
\begin{figure}[!htb]
		\centering
		\includegraphics[width=\textwidth]{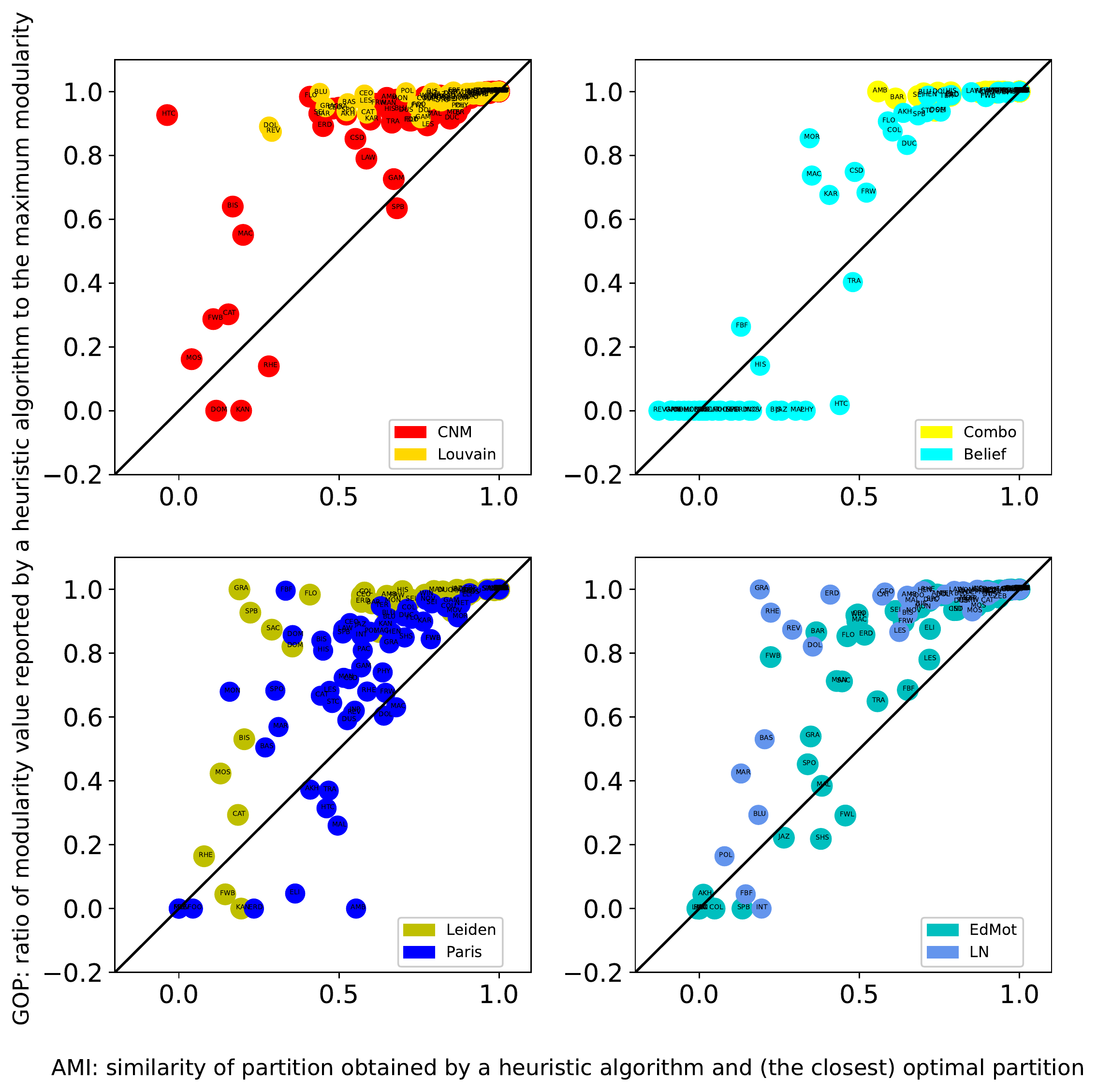}
		\caption{Global optimality percentage and normalized adjusted mutual information measured for eight modularity maximization heuristics in comparison with (all) globally optimal partitions. (Magnify the high-resolution figure on screen for more details.)}
		\label{fig:heuristic}
	\end{figure}

	For summarizing the results of eight heuristics on 80 networks, we present four scatter-plots of GOP and AMI. Figure \ref{fig:heuristic} shows GOP on the y-axes and AMI on the x-axes for the combination of each network and algorithm. For each algorithm (color-coded), there are 60 data points for the 60 real networks and 2 data points representing the average of 10 Erd\H{o}s-R\'{e}nyi and the average of 10 Barab\'{a}si-Albert graphs. The first three letters of the network names are indicated on each data point (magnify the figure on screen for the details). Four 45-degree lines are drawn to indicate the areas where the GOP and AMI are equal. In other words, the 45-degree lines show areas where the extent of sub-optimality (1-GOP) is associated with a dissimilarity (1-AMI) of the same size between the sub-optimal partition and any optimal partition.

	Looking at the y-axes values in Figure \ref{fig:heuristic}, we observe that there is a substantial variation in the values of GOP (i.e.\, the extent of sub-optimality) for the eight heuristic algorithms. The Belief algorithm returns partitions associated with negative modularity values for 45 of the 80 instances (leading to most of its data points having GOP$=0$ and being concentrated at the bottom of the scatter-plot). The Paris algorithm returns partitions with modularity values substantially smaller than the maximum modularity values. Aside from a few exceptions, all data points for Leiden and LN have the same position indicating their identical performance on most of these instances. The two algorithms CNM and EdMot seem to have higher variation in GOP (compared to the other algorithms) for these instances. Overall, the four algorithms with highest and increasing performance in returning close-to-maximum modularity values are LN, Leiden, Louvain, and Combo respectively. Despite that these instance are graphs with no more than 2812 edges, they are, according to Figure \ref{fig:heuristic}, challenging instances for these heuristic algorithms to optimize. Given that modularity maximization is an NP-complete problem \cite{brandes2007modularity,meeks2020parameterised}, one can argue that the performance of these heuristic methods in term of proximity to an optimal partition does not improve for larger networks.
	
	The x-axes values in Figure \ref{fig:heuristic} show considerable dissimilarity between the sub-optimal partitions and an optimal partition for these 80 instances. Except for the Combo algorithm, a large number of the sub-optimal partitions obtained by these heuristic algorithms have AMI values smaller than {$0.6$}. This indicates that their sub-optimal partitions are substantially different from any optimal partition. Even for data points concentrated at the top of the scatter-plots which have $0.95<$GOP$<1$, we see AMI values substantially smaller than 1. Compared to the other seven heuristics, Combo appears to consistently return partitions with large AMIs on a larger number of these 80 instances. 
	
	Focusing on the position of data points, we observe that they are mostly located above their corresponding 45-degree line. This indicates that sub-optimal partitions tend to be disproportionately dissimilar to any optimal partition (as foreshadowed in \cite{dinh_network_2015}). This result goes against the naive viewpoint that close-to-maximum modularity partitions are also close to an optimal partition. Our results are aligned with previous concerns that these heuristics may result in degenerate solutions far from the underlying community structure \cite{good_performance_2010} and they have a high risk of algorithmic failure \cite{kawamoto2019counting}. 
	
	\subsection{Success rate of heuristic algorithms in maximizing modularity}\label{sec:results4}
	
	Our GOP results for the eight heuristic algorithms allow us to answer a fundamental question about the heuristic modularity maximization algorithms: how often each algorithm returns an optimal (a maximum-modularity) partition? We report the fraction of networks (out of 80) for which a given algorithm returns an optimal partition. Combo \cite{sobolevsky2014general} has the highest success rate, returning an optimal partition for $55\%$ of the networks. LN \cite{rb_pots_2008} and Leiden \cite{traag_louvain_2019} maximize modularity for 36.2\% of the networks considered. Louvain \cite{blondel_fast_2008} has a success rate of 18.7\%. The algorithms CNM \cite{clauset_finding_2004}, EdMot \cite{edmot_2019}, Paris \cite{paris_2018}, and Belief \cite{zhang2014} have success rates of 5\%, 2.5\%, 1.2\%, and 0\% respectively. These are arguably low success rates for what the name \textit{modularity maximization algorithm} implies or the idea of discovering network communities through maximizing a function.
	
	Earlier in Figure \ref{fig:heuristic}, we observed that near-optimal partitions tend to be disproportionately dissimilar to any optimal partition. In other words, close-to-maximum modularity partitions are rarely close to any optimal partition. Taken together with the low success rates of heuristic algorithms in maximizing modularity, our results indicate a crucial mismatch between the design philosophy of modularity maximization algorithms for CD and their capabilities: heuristic modularity maximization algorithms rarely return an optimal partition or a partition resembling an optimal partition.
	
	\section{Discussions and Future Directions}
	
	Understanding modularity capabilities and limitations has been complicated by the under-studied sub-optimality of modularity-based heuristics and their methodological consequences. Previous methodological studies \cite{lancichinetti_limits_2011,Chen_2018,global2018,Miasnikof2020,peixoto_2023}, which have shed light on other aspects, have rarely disentangled the heuristic aspect of these algorithms from the fundamental concept of modularity. Our study is a continuation of previous efforts \cite{good_performance_2010} in separating the effects of sub-optimality (or the choice of using greedy algorithms \cite{kawamoto2019counting}) from the effects of using modularity on the fundamental task of detecting communities. 
	
	We analyzed the effectiveness of eight heuristics in maximizing modularity. While our findings are limited to a few algorithms, their combined usage by tens of thousands of peer-reviewed studies \cite{Kosowski2020} motivates the importance of this assessment. Most heuristic algorithms for modularity maximization tend to scale well for large networks \cite{zhao2021community}. They are widely used not only because of their scalability or ease of implementation \cite{kawamoto2019counting}, but also because their high risk of algorithmic failure is not well understood \cite{kawamoto2019counting}. The scalability of these heuristics comes at a cost: their partitions have no guarantee of proximity to an optimal partition \cite{good_performance_2010} and, as our results showed, they rarely return an optimal partition. Moreover, we showed that their sub-optimal partitions tend to be disproportionately dissimilar to any optimal partition. 
	
	Neither using modularity nor succeeding in maximizing it is required for CD at the big-picture level. A recent study suggests modularity maximization is the most problematic CD method and considers it harmful \cite{peixoto_2023}. Another study shows that, given computational feasibility, exact maximization of multiresolution modularity outperforms other CD methods in accurate and stable retrieval of planted communities \cite{aref2022bayan} suggesting the relevance of modularity for CD. For some applications and contexts, \textit{general} CD algorithms \cite{peel2017ground} which scale to large instance sizes are needed. However, for a ``narrow set of tasks'' \cite[pp.7]{peel2017ground}, involving small and mid-sized networks, \textit{specialized} algorithms which outperform general algorithms are useful. 
	
	Our findings suggest that if modularity is to be used for detecting communities, developing approximation \cite{cafieri2014reformulation,dinh_network_2015,kawase2021} and exact \cite{aloise_column_2010,aref2022bayan} algorithms are recommendable for a more methodologically sound usage of modularity within its applicability limits. 
	Exact algorithms can also reveal the formal guarantees of performance \cite{fortunato2022newman} for accurate modularity-based algorithms.
	
	A promising path forward could be using the advances in integer programming to develop a specialized accurate algorithm for solving the modularity maximization IP models \cite{brandes2007modularity,agarwal_modularity-maximizing_2008,dinh_toward_2015} for networks of practical relevance within the limits of computational feasibility.
	New heuristic and approximation algorithms that strike a balance between accurate computations and scalability may also be useful particularly for large-scale networks.
	
	\subsubsection{Author contributions}
	Conceptualization (SA); data curation (SA, HC); formal analysis (SA, MM); funding acquisition (SA); investigation (SA); methodology (SA,MM); project administration (SA); resources (SA, MM); software (SA, HC, MM); supervision (SA, MM); validation (SA, MM); visualization (SA, MM); writing - original draft preparation (SA); writing - review \& editing (SA, MM).
	
	\subsubsection{Acknowledgements} We are thankful to the three anonymous referees and Ly Dinh for their helpful comments. We acknowledge Zachary P. Neal for pointing us to this problem and Santo Fortunato for the encouraging correspondence. Accessing  \href{https://cdlib.readthedocs.io/}{CDlib}, \href{https://networks.skewed.de/}{Netzschleuder}, and \href{https://icon.colorado.edu/}{ICON} has been particularly helpful in this study for which we thank
	Giulio Rossetti, Tiago P. Peixoto, Aaron Clauset, and everyone contributing to these open science initiatives. This study has been supported by the Data Sciences Institute at the University of Toronto.

	\bibliographystyle{splncs04}
	\bibliography{refs}

	\section*{Appendix}
	
	The data on 20 random graphs used in this study are available in a \textit{FigShare} data repository \cite{Aref2023figsharenonote}. The 60 real networks are loaded as simple undirected graphs. They are available in the publicly accessible network repository \href{https://networks.skewed.de/}{Netzschleuder} with the 60 names below:
	
	dom, packet\_delays, sa\_companies, ambassador, florentine\_families, rhesus\_monkey, kangaroo, internet\_top\_pop, high\_tech\_company, moviegalaxies, {november17}, moreno\_taro, sp\_baboons, bison, dutch\_school, zebras, cattle, moreno\_sheep, 7th\_graders, college\_freshmen, hens, freshmen, karate, dutch\_criticism, montreal, ceo\_club, windsurfers, elite, macaque\_neural, sp\_kenyan\_households, contiguous\_usa, cs\_department, dolphins, macaques, terrorists\_911, train\_terrorists, highschool, law\_firm, baseball, blumenau\_drug, lesmis, fresh\_webs, sp\_office, swingers, polbooks, game\_thrones, football, football\_tsevans, sp\_high\_school\_new, foodweb\_baywet, revolution, foodweb\_little\_rock, student\_cooperation, jazz\_collab, interactome\_pdz, physician\_trust, malaria\_genes, marvel\_partnerships, facebook\_friends, netscience
	
	For more information on each network and its original source, one may check the Netzschleuder website by adding the network name at the end of the url:  https://networks.skewed.de/net/. For example, \url{https://networks.skewed.de/net/malaria_genes} provides additional information for the \textit{malaria\_genes} network. In cases of multiple networks existing with the same name in Netzschleuder, we have used the lexicographically first network (e.g. we have used the \textit{HVR\_1} network from \url{https://networks.skewed.de/net/malaria_genes}).
	
\end{document}